\documentclass{aa}
\usepackage{graphicx}
\usepackage{natbib}
\usepackage{color}
\newcommand{\tw}{\color{black}}
\usepackage{txfonts}
\begin{document}
\title{Optimization approach for the computation of magnetohydrostatic
coronal equilibria in spherical geometry}
   \subtitle{}

   \author{T. Wiegelmann \inst{1}
           \and
          T. Neukirch \inst{2}
          \and
          P. Ruan \inst{1}
          \and
          B. Inhester \inst{1}
          }

   \offprints{T. Wiegelmann}

   \institute{Max-Planck-Institut f\"ur Sonnensystemforschung,
Max-Planck-Strasse 2, 37191 Katlenburg-Lindau, Germany\\
              \email{wiegelmann@mps.mpg.de}
     \and
       School of Mathematics and Statistics,
   University of St. Andrews,
   St. Andrews, KY16 9SS,
   United Kingdom \\
             }

   \date{}


  \abstract
  {This paper presents a method which can be used to calculate models of the global solar corona from observational data.} 
 %
   {We present an optimization method for computing nonlinear
   magnetohydrostatic equilibria in spherical geometry with the aim to obtain
   self-consistent solutions for the coronal magnetic field, the coronal plasma density and
   plasma pressure using observational data as input.}
  {
 Our code for the self-consistent computation of the coronal magnetic
 fields and the coronal plasma solves the non-force-free magnetohydrostatic
 equilibria using an optimization method.
 Previous versions of the code have been used to
 compute non-linear force-free coronal magnetic fields from photospheric measurements in
 Cartesian and spherical geometry, and magnetostatic-equilibria in Cartesian geometry.
 We test our code with the help of a known analytic 3D equilibrium solution of the magnetohydrostatic equations. The detailed comparison between the numerical calculations and the exact equilibrium solutions is made by using magnetic field line plots, plots of
 density and pressure and some of the usual quantitative numerical comparison measures.}
   {
   {We find that the method reconstructs the equilibrium accurately,
with residual forces of the order of the discretisation error of the analytic
solution. The correlation with the reference solution is better than
$99.9\%$ and the magnetic energy is computed accurately  with an
error of $< 0.1 \%$.
}
   }
   {
   We applied the method so far to an analytic test case.
   We are planning to use this method with real observational data as input as soon as possible.
    }

 \authorrunning{Wiegelmann et al.}
 \titlerunning{Spherical optimization code}
\maketitle
%

\section{Introduction}
In the recent past, numerical methods based on optimization principles have been used for a number of problems associated
with the calculation of solar MHD equilibria. \citet{wheatland:etal00} were the first to suggest the use of an optimization
method to calculate nonlinear force-free fields in the corona from photospheric measurements. Since then the optimization
method has been extended in various ways,
for example by improving certain aspects of the original method for force-free fields
\citep[e.g.][]{wiegelmann04}, by introducing additional features such as plasma pressure into the
method \citep[e.g.][]{wiegelmann:etal06b} or by
reformulating the method in other geometries  \citep[e.g.][]{wiegelmann07}.

Optimization methods have the advantage of being conceptually straightforward and are reasonably
easy to implement \citep[see e.g.]{wheatland:etal00,wiegelmann:etal03,inhester:etal06}. At the moment they also seem to
be very competitive in terms of computational efficiency \citep[e.g][]{schrijver:etal06,metcalf:etal07}.
A slight disadvantage compared to, for example, the Grad-Rubin method
\citep[e.g.][]{amari:etal97,wheatland06,inhester:etal06,amari:etal06} is that they still lack the same degree
of rigorous mathematical basis existing for other methods.

In the present paper we describe a further extension of the optimization method to calculate
magnetohydrostatic (MHS) equilibria (including pressure and gravity) in spherical geometry.
This is important for calculating global models of the corona including information going beyond
just the structure of the magnetic field. In section \ref{equations} we describe the basic equations
of the optimization method for problem in hand. We then give a brief description of the analytical
3D MHS equilibria that we use to test the numerical code in section \ref{3DMHS} and present the
test results in section \ref{results}. Our conclusions are presented in section \ref{conclusions}.


\section[]{Basic equations}
\label{equations}
The magnetohydrostatic (MHS) equations are given by
\begin{eqnarray}
(\nabla \times {\bf B}) \times {\bf B}
-\mu_0\nabla p -\mu_0 \, \rho \, \nabla \Psi &=& {\mathbf{0}}  \label{1} \\
\nabla \cdot {\bf B} &=& 0,  \label{2}
\end{eqnarray}
where ${\bf B}$ is the magnetic field, $p$ the plasma pressure, $\rho$ the mass density
and $\Psi=-\frac{G M_s}{r}$ the gravitational potential with the gravitational constant $G$, the
solar mass $M_s$ and the distance from the sun's center $r$. We do not assume an equation
of state for the coronal plasma, but leave $p$ and $\rho$ to be independent quantities.
To find a magnetic field $\mathbf{B}$, plasma pressure $p$ and plasma density $\rho$ satisfying
Eqs. (\ref{1}) and (\ref{2}), we follow the spirit of the previous optimization methods
\citep[e.g.][]{wheatland:etal00,wiegelmann:etal03a,wiegelmann04,wiegelmann:etal06b} and define the functional
\begin{eqnarray}
L({\bf B},p,\rho) & = & \int \Bigg[\frac{\left|(\nabla \times {\bf B}) \times {\bf B}
-\mu_0 \, \nabla p -\mu_0 \, \rho \,\nabla \Psi    \right|^2}{B^2} \nonumber \\
&& \; + \; |\nabla \cdot {\bf B}|^2\Bigg] \; \, r^2 \, \sin \theta \, dr \, d \theta \, d \phi.
\label{defL_Bprho}
\end{eqnarray}

It is obvious that
Eqs. (\ref{1}) and (\ref{2}) are satisfied if $L=0$.
Here ${\bf B}$ is a vector field, but not necessarily a solenoidal magnetic field during
the iteration.
The numerical method is based on an iterative scheme to minimize the functional $L$. To simplify the
mathematical derivation
we define the quantities
\begin{eqnarray}
{\bf \Omega_a} &=& \; B^{-2} \;\left[(\nabla \times {\bf B})
\times {\bf B} -\mu_0  \, \nabla p -\mu_0 \,\rho\, \nabla \Psi
\right] \\
{\bf \Omega_b} &=& B^{-2} \;\left[(\nabla \cdot {\bf B}) \; {\bf B} \right],
\label{defomega}
\end{eqnarray}
and rewrite $L$ as
\begin{equation}
L=\int_{V} \;  \left[ B^2 \Omega_a^2+  B^2 \Omega_b^2 \; r^2 \right] \, \sin \theta \, dr \, d \theta \, d \phi.
\end{equation}
Taking the derivative of $L$  with respect to an iteration
parameter $t$, where ${\bf B}, \; p, \; \rho$ are assumed to depend upon $t$, we obtain
\begin{eqnarray}
\frac{1}{2} \; \frac{d L}{d t} &=&-\int_{V} \frac{\partial {\bf
B}}{\partial t} \cdot {\bf F} \; dV
+\int_{V} \frac{\partial p}{\partial t} \; \mu_0 \nabla \cdot {\bf \Omega_a} \; dV \nonumber \\
&&-\int_{V} \frac{\partial \rho}{\partial t} \; \mu_0 {\bf \Omega_a} \cdot \nabla \Psi \; dV
-\int_{S} \frac{\partial {\bf B}}{\partial t} \cdot {\bf G} \; dS \nonumber \\
&& -\int_{S} \frac{\partial p}{\partial t} \; \mu_0 {\bf \Omega_a} \cdot {\bf dS},
\end{eqnarray}
where
\begin{eqnarray}
{\bf F} & =& \nabla \times ({\bf \Omega_a} \times {\bf B} ) - {\bf \Omega_a}
\times (\nabla \times {\bf B}) \nonumber \\
&& + \nabla({\bf \Omega_b} \cdot {\bf B})-  {\bf \Omega_b}(\nabla \cdot {\bf B})
+(\Omega_a^2 + \Omega_b^2) \, {\bf B}
\end{eqnarray}
and
\begin{eqnarray}
{\bf G} & = & {\bf \hat n} \times ({\bf \Omega_a} \times {\bf B} )
-{\bf \hat n} (\bf \Omega_b \cdot \bf B).
\end{eqnarray}
Assuming that
\begin{eqnarray}
\frac{\partial {\bf B}}{\partial t} &=& \mu \; {\bf F} \label{Beq}\\
\frac{\partial p}{\partial t} &=& -\nu \; \mu_0 \nabla \cdot {\bf \Omega_a} \label{peq} \\
\frac{\partial \rho}{\partial t} &=& \xi \; \mu_0 \; {\bf \Omega_a} \cdot \nabla \Psi
=\xi \; \mu_0 \frac{G M_s}{r^2} \; {\bf \Omega_a} \cdot {\bf e_r} \label{rhoeq}
\end{eqnarray}
with positive definite parameters $\mu$, $\nu$ and $\xi$ and that the boundary integrals vanish, one can easily see that
 $L$ is monotonically decreasing with $t$ (note that this does not necessarily imply that $L$ tends to zero).

Discretized versions of Eqs. (\ref{Beq}) to (\ref{rhoeq}), together with appropriate boundary conditions, form the basis for the numerical scheme.
The boundary conditions have to be consistent with the assumption that the boundary integrals vanish, for example, by keeping the magnetic field, pressure and density fixed on the boundaries during the iteration. For testing the method in this paper we shall take these boundary conditions from the known exact solution. For practical applications these would have to come from observational data. We remark that due to the introduction of additional forces the constraints on the consistency of the boundary conditions for the magnetic field are somewhat different from the force-free case. However, the general theory of magnetohydrostatic equilibria requires for example that the pressure has the same value at both foot points of a closed field line under the general conditions assumed in the present paper. This is similar to the Cartesian case discussed by \citet{wiegelmann:etal06b}, where the pressure equation was forward integrated along field lines using an upwind method from one foot point to the other (in the test case  described later we make use of the property that the pressure is known to be consistent).

In the numerical implementation based on this method one has to choose the product of the time-step $\Delta t$ with the three numerical parameters $\mu$, $\nu$ and $\xi$. Usually these products have to be small enough to achieve convergence and this is ensured by an adaptive time-step control. In this paper we have chosen the three parameters (multiplied by $\Delta t$) to have the same values on all grid points. Previous experience with applying a similar method to force-free magnetic fields in spherical geometry \citep{wiegelmann07} showed that choosing the same values for the entire box can lead to  long computing times in the
polar regions due to the distortion of the numerical grid in spherical polar coordinates  towards the poles (note that the poles $\theta=0$ and $\theta=\pi$ are excluded from the computational domain). This could in principle be compensated by allowing for a spatial variation of the iteration parameters.

\section{3D MHS Equilibria}
\label{3DMHS}

We use the exact 3D MHS equilibrium in spherical coordinates presented by \citet{neukirch95} to test our code (called case II in his paper).
In his paper, \citet{neukirch95}  extended earlier work by \citet{bogdan:low86} on exact 3D MHS equilibria in
spherical coordinates.
The general method was first found by \citet{low85} (previous closely related work can
also be found in \citet{low82} and \citet{low85}) for Cartesian coordinates and developed further in
\citet{low91,low92,low93a,low93b,low05}. The method relies on the presence of an external force (in our case gravitation)
and the basic assumption that the electric currents flow only in surface direction perpendicular to the direction of the
external force. The additional assumption that the dependence of the current density on the spatial coordinates has a special
form involving the magnetic field component along the direction of the external force (in our case that is the radial component
of $\mathbf{B}$ leads to a linear equation for that magnetic field component. The plasma density and pressure are determined from
the force balance equation.

\citet{neukirch95} has extended this method by including a current density component of the constant-$\alpha$
type ($\alpha \mathbf{B}$ with $\alpha$ constant), which also allows
components of the current density in the
direction of the external force.  It is important to emphasize that this does not mean that the
{\em total} field-aligned
current density is of the linear force-free type, because the other component of the current density also has a field-aligned part.

The formulation by \citet{neukirch95}
basically reduces the mathematical problem
to the solution of an equation similar to a Schr\"odinger equation.
A slightly simpler formulation of the same
problem was given by \citet{neukirch:rastatter99} and some analytical
solutions (in Cartesian coordinates) with a nonlinear relationship
between the current density and the magnetic field were found by \citet{neukirch97}. A formulation using Green's functions was given by \citet{petrie:etal00}.

\begin{figure}
\includegraphics[bb=80 80 320 370,clip,height=6cm,width=5.5cm]{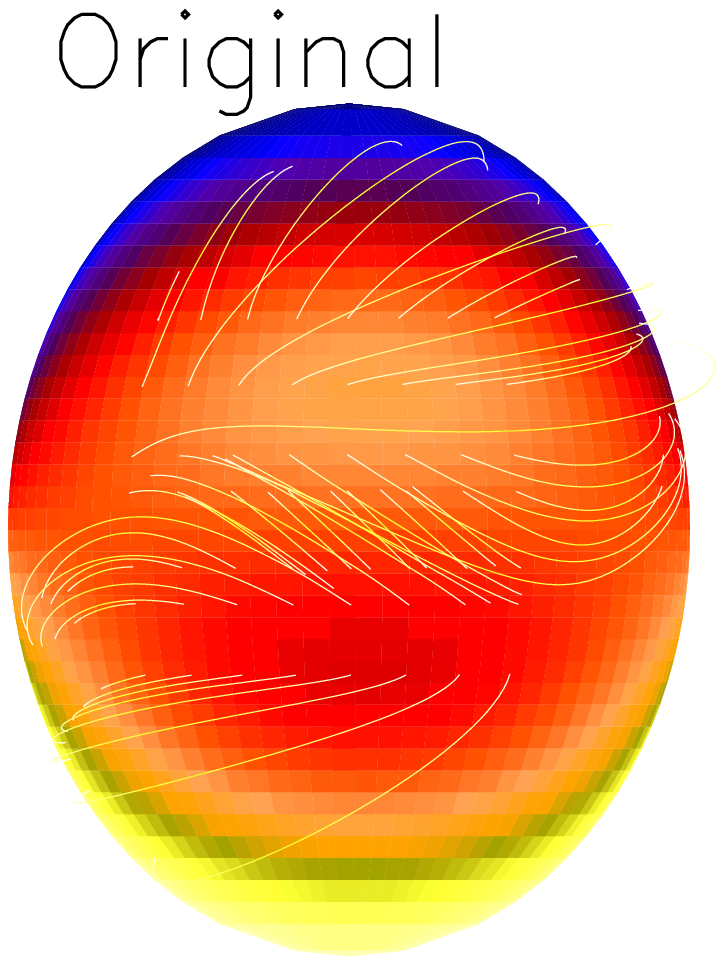}
\includegraphics[bb=80 80 320 370,clip,height=6cm,width=5.5cm]{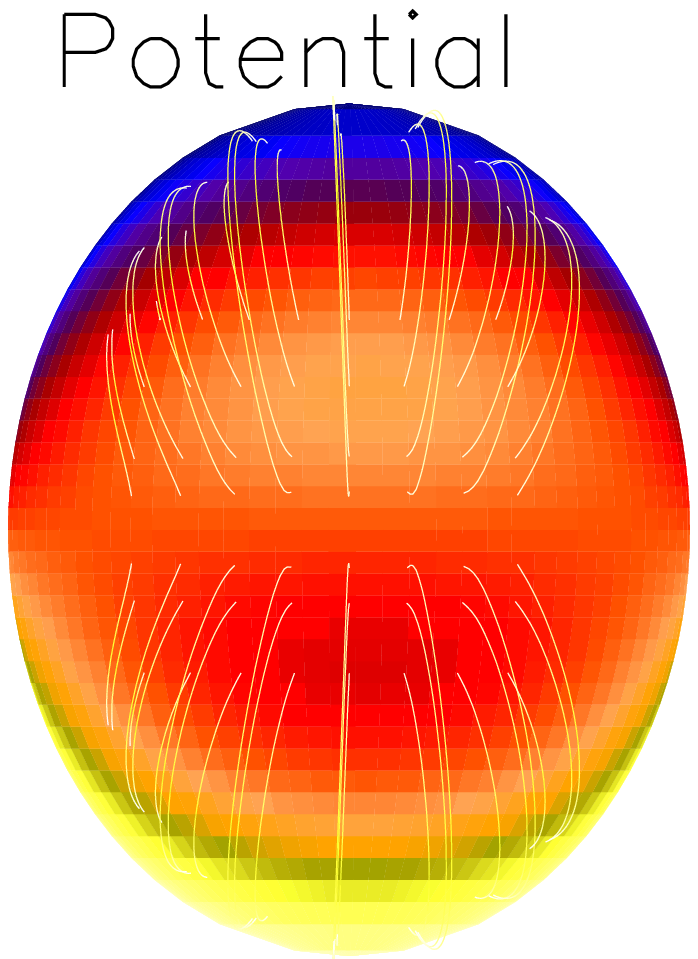}
\includegraphics[bb=80 80 320 370,clip,height=6cm,width=5.5cm]{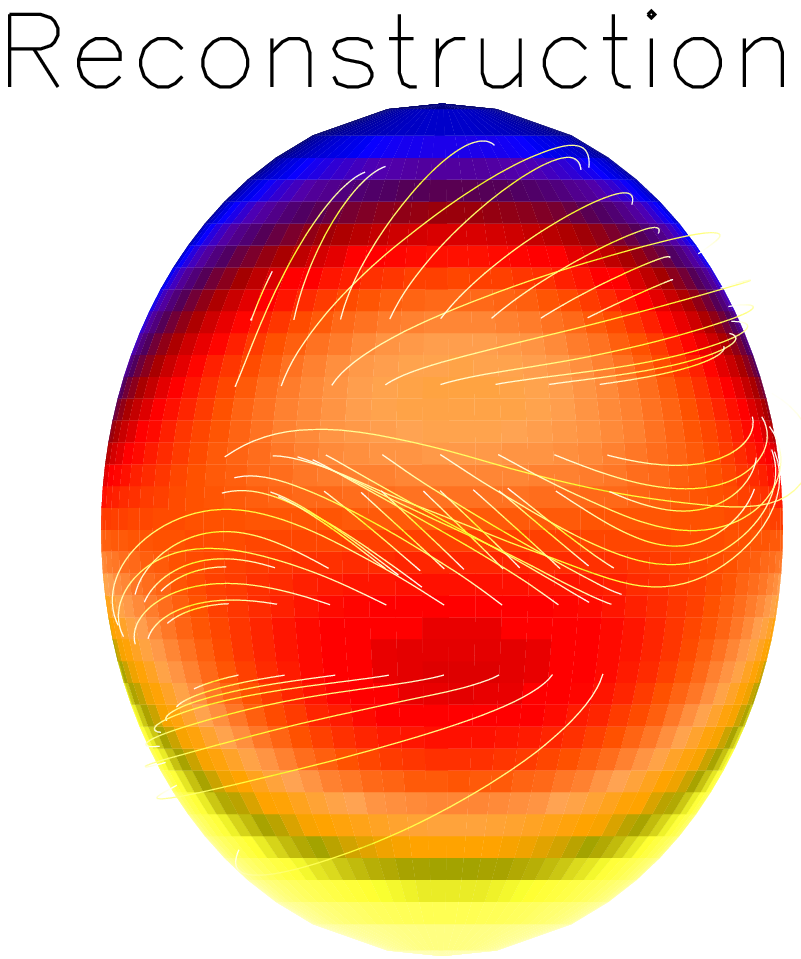}
\caption{Magnetic field lines for the exact analytical solution, a potential field with the same $B_r$ on the
photospheric boundary ($r=R_{\sun}$)
and the magnetic field obtained by the optimization method.
The color coding corresponds to the value of $B_r$ on the photosphere
(yellow:positive, blue: negative) and
the disk center corresponds to $\phi=180^\circ$. The potential field is used as the initial
field for the numerical calculation and clearly has a different connectivity from the exact
analytical solution. The reconstructed solution matches the analytical solution
down to plotting precision, except for one equatorial field line.
}
\label{fig1}
\end{figure}

Since the magnetic field given in Eqs. (45)-(47) of \citet{neukirch95} contain a number of typographical errors, we repeat the correct field components here. As in \citet{neukirch95} we define the functions
\begin{eqnarray}
f_1(r) &=& \frac{\cos q +q\sin q}{\cos q_0 +q_0\sin q_0} \\
f_2(r) &=& \frac{(3-q^2)\cos q+3 q\sin q}{(3-q_0^2)\cos q_0+3 q_0\sin q_0}
\end{eqnarray}
with
\begin{equation}
q =\alpha(r+a),\quad q_0=\alpha(r_0+a).
\end{equation}
We then obtain\footnote{Compared to \citet{neukirch95} these expression have been corrected in the following way: a) a factor $1/2$ has been included in the second term of $B_r$, b) a factor $1/\sin\theta$ has been removed from the second term of $B_\phi$ and c) the sign of the third term of $B_\phi$ is negative.}
\begin{eqnarray}
B_r &= & A_{10} \frac{r_0^2}{r^2}\frac{r_0+a}{r+a} f_1(r) Y_1^0 - \nonumber \\
 & & \mbox{\hspace{12pt}} \frac{1}{2} A_{21} \frac{r_0^2}{r^2}\frac{(r_0+a)^2}{(r+a)^2} f_2(r)
        (Y_2^1-Y_2^{-1})\\
B_\theta &=& \frac{1}{2} A_{10}\frac{r_0^2}{r}\frac{r_0+a}{r+a} \left(\frac{d f_1}{dr}-\frac{f_1}{r+a}\right)
                  \frac{\partial}{\partial \theta} Y_1^0 \nonumber \\
       & & -\frac{1}{12} A_{21}\frac{r_0^2}{r}\frac{(r_0+a)^2}{(r+a)^2}
        \left(\frac{d f_2}{dr}-\frac{2 f_2}{r+a}\right)
                  \frac{\partial}{\partial \theta} (Y_2^1  - Y_2^{-1}) \nonumber\\
      & & -\frac{1}{12} A_{21}\alpha\frac{r_0^2}{r}\frac{(r_0+a)^2}{(r+a)^2} f_2(r) \frac{1}{\sin\theta}
      \frac{\partial}{\partial \phi}(Y_2^1-Y_2^{-1}) \\
B_\phi & =& -\frac{1}{2} A_{10} \alpha \frac{r_0^2}{r} \frac{r_0+a}{r+a} f_1(r) \frac{\partial}{\partial \theta} Y_1^0 \nonumber \\
& & + \frac{1}{12} A_{21} \frac{r_0^2}{r}\frac{(r_0+a)^2}{(r+a)^2}
         f_2(r)\frac{\partial}{\partial \theta} (Y_2^1  - Y_2^{-1}) \nonumber\\
& & \mbox{\hspace{-14pt}}- \frac{1}{12} A_{21} \frac{r_0^2}{r}\frac{(r_0+a)^2}{(r+a)^2}
        \left(\frac{d f_2}{dr}-\frac{2 f_2}{r+a}\right)\frac{1}{\sin\theta}\frac{\partial}{\partial \phi} (Y_2^1-Y_2^{-1}).
\end{eqnarray}
These expression tend to the solution presented as case III in \citet{bogdan:low86} in the limit $\alpha =0$. The
magnetic field is shown in the top panel of Fig. \ref{fig1}.
The solution is completed by expressions for the pressure and the density as given by
 \citet{bogdan:low86} and \citet{neukirch95}.

\citet{neukirch95} followed \citet{bogdan:low86} and normalized the radial coordinate $r$ at a radius of  $1.5 \, R_{\sun}$.
In the present paper, we deviate from this and normalize the radial coordinate at a radius of $1 \,R_{\sun}$, because we want
to impose our boundary conditions at the solar surface ($=1 \,R_{\sun}$). We then carry out
our numerical calculation on a box which has an inner boundary at
$1 \,R_{\sun}$ and  an outer boundary at $ (1.0+\pi/2) \,R_{\sun} \approx 2.57 \,R_{\sun} $.
We use a grid with $20$ points
in the radial direction, $40$ points in the latitudinal direction and $80$ points in the  longitudinal direction.
We exclude the polar regions for numerical reasons (see discussion in section \ref{equations}) and  extend the
numerical box in $\theta$ only from
$11.25^{\circ}$ to $168.75^{\circ}$.

As parameters we choose the parameter $a$, which determines the influence of the
non-magnetic forces on the equilibrium, to have the value
$a=0.2$, and we choose the force-free parameter $\alpha$ to have the value $\alpha=0.5$.
Internally our code normalizes the length scale with
one solar radius, the magnetic field strength with the average
radial magnetic field strength on the photosphere $B_{\rm ave}$,
the pressure with ${\tilde p }=\mu_0 p/B_{\rm ave}^2$
and the mass density with ${\tilde \rho}=\mu_0 \, G \, M_s \rho/R_s \, B_{\rm ave}^2$
\begin{figure}
\mbox{
\includegraphics[bb=0 0 410 410,clip,height=3cm,width=4.0cm]{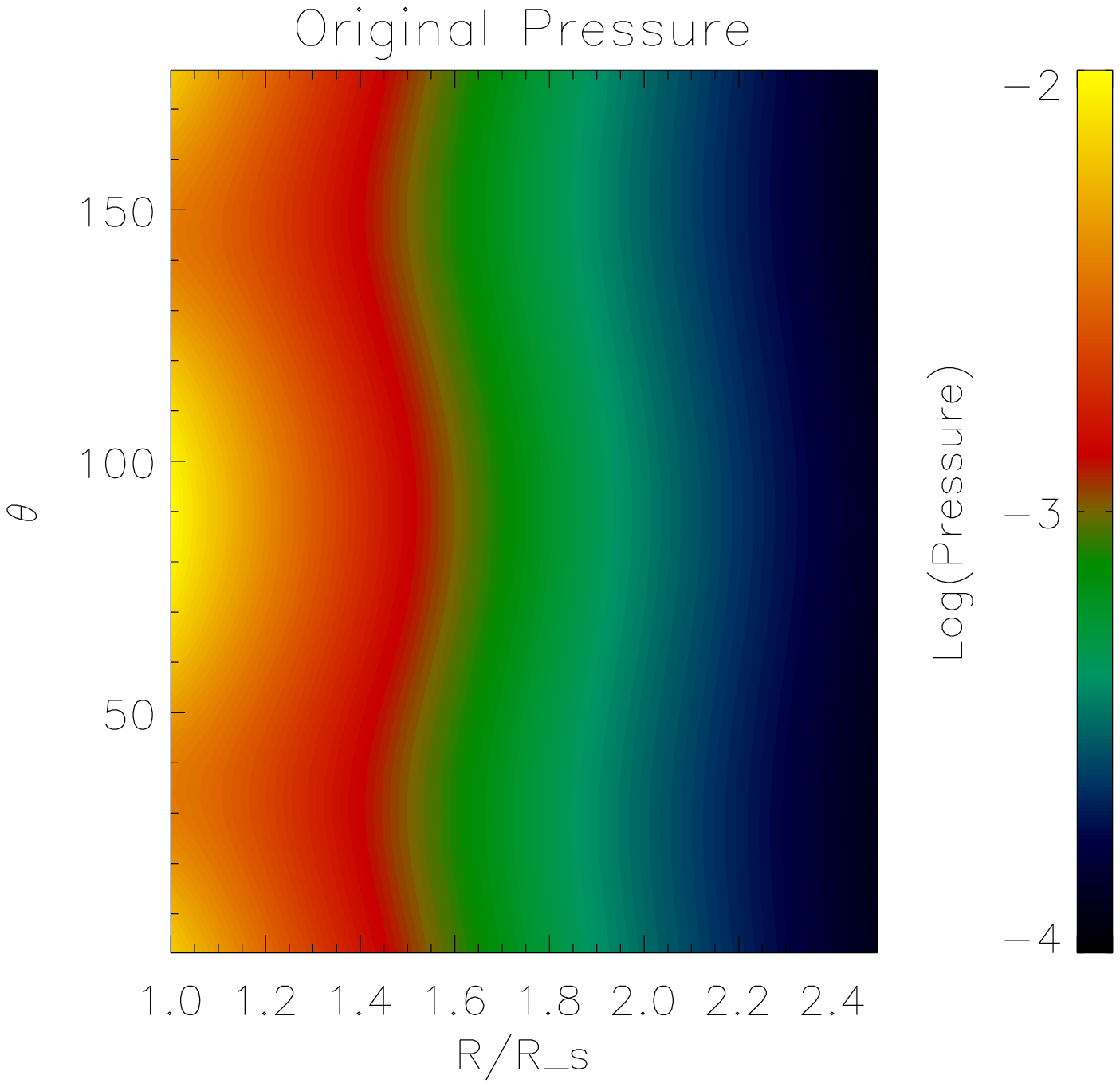}
\includegraphics[bb=0 0 410 410,clip,height=3cm,width=4.0cm]{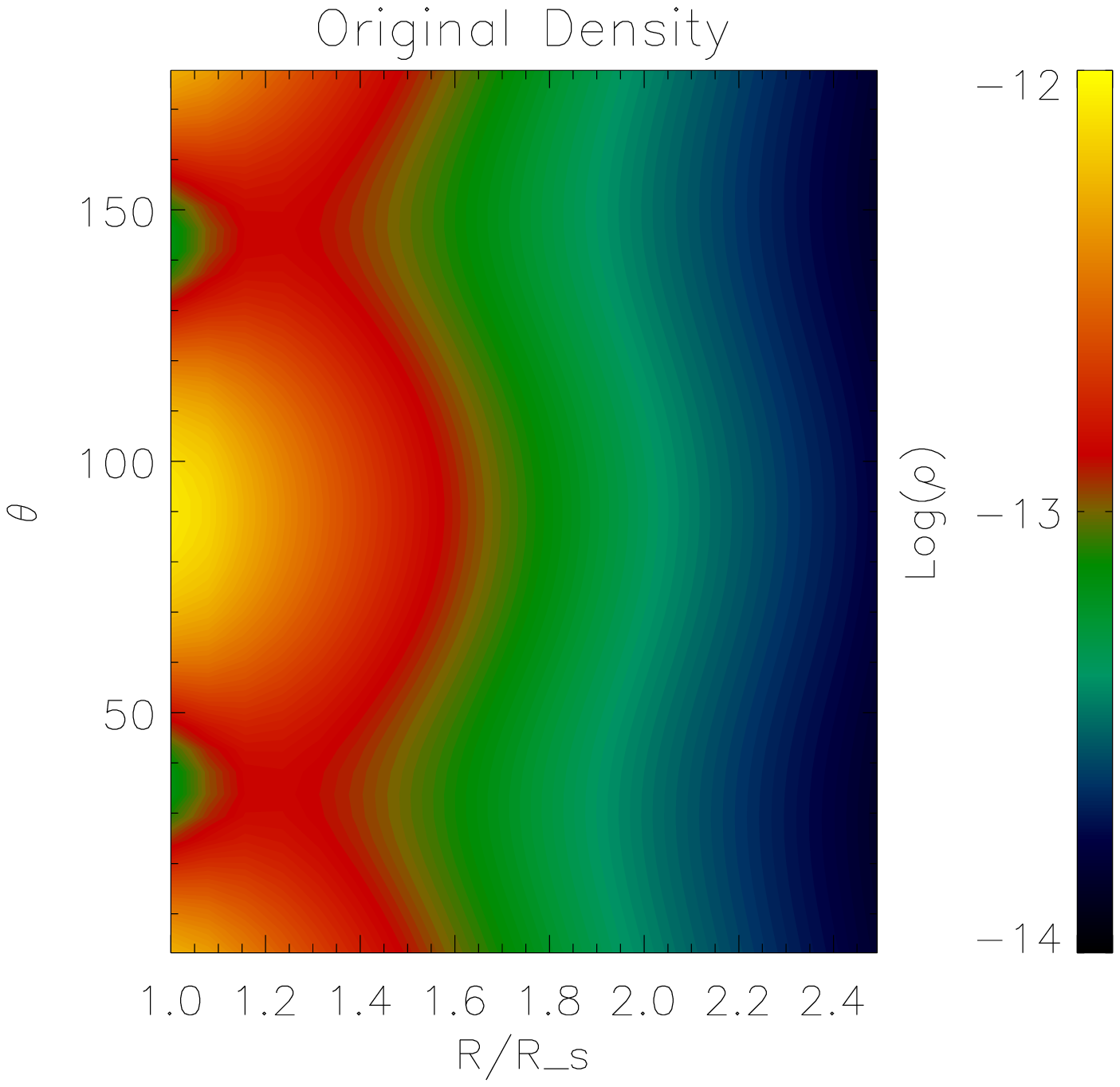}}
\mbox{
\includegraphics[bb=0 0 410 410,,clip,height=3cm,width=4.0cm]{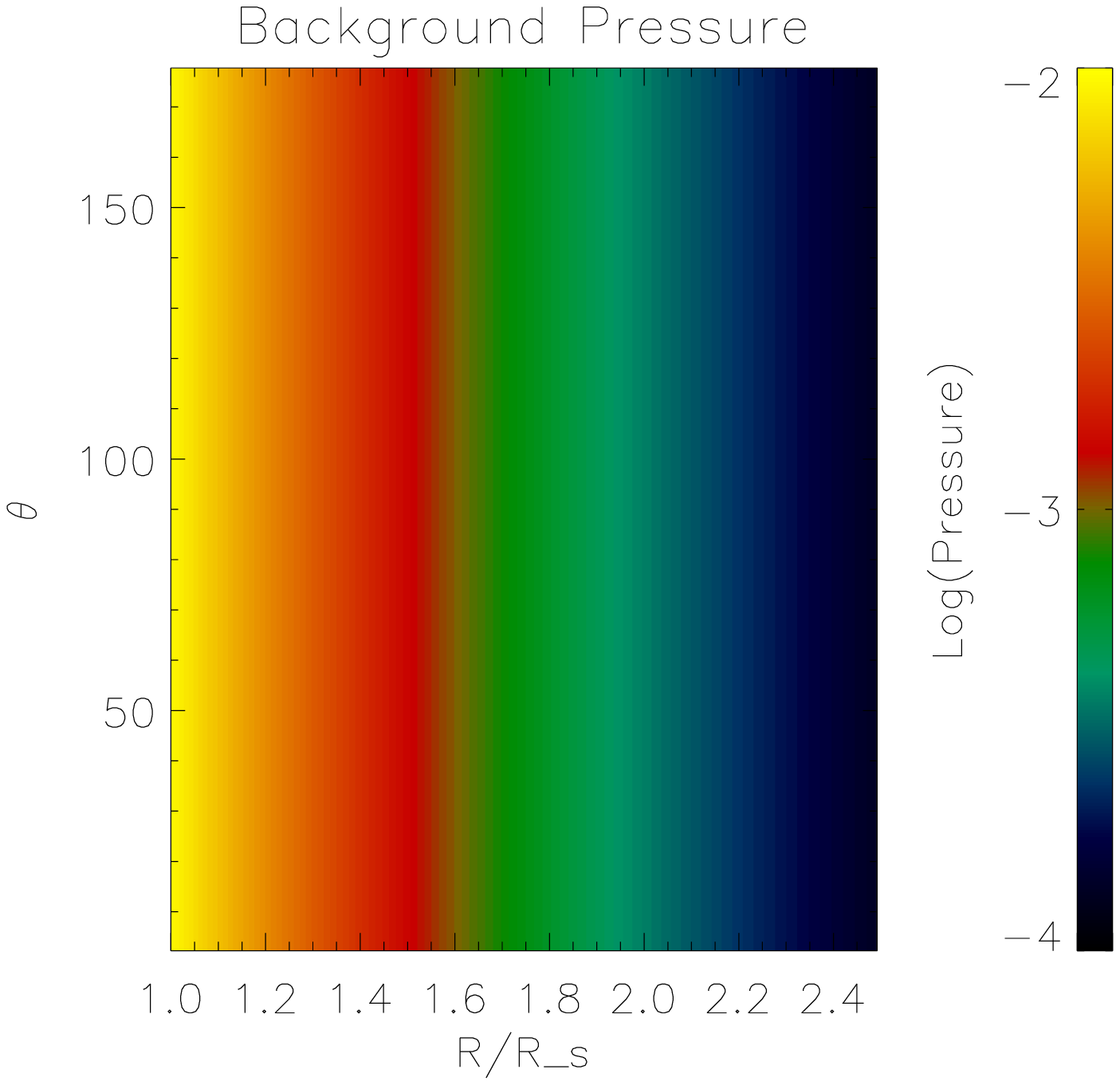}
\includegraphics[bb=0 0 410 410,,clip,height=3cm,width=4.0cm]{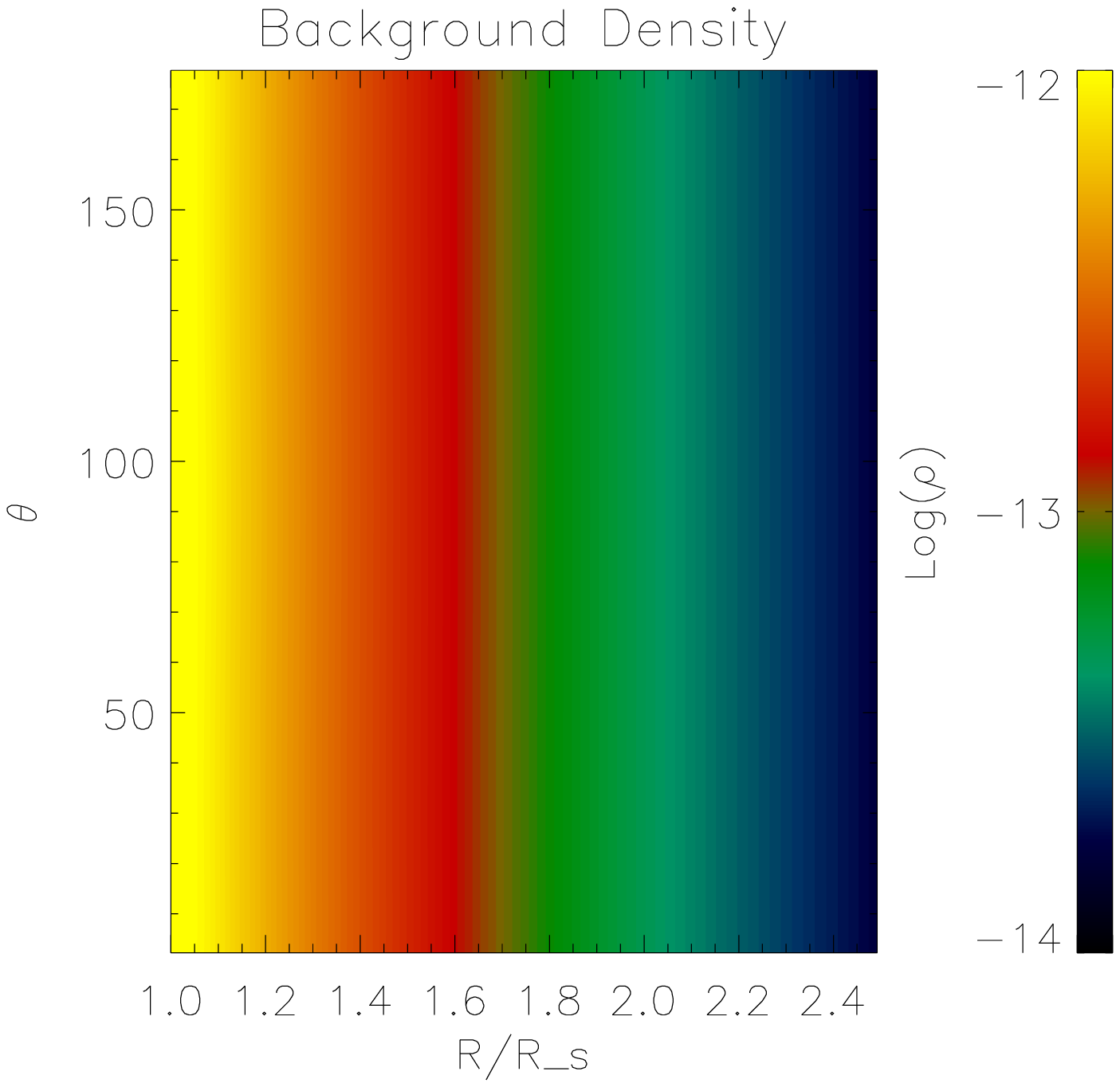}}
\mbox{
\includegraphics[bb=0 0 410 410,,clip,height=3cm,width=4.0cm]{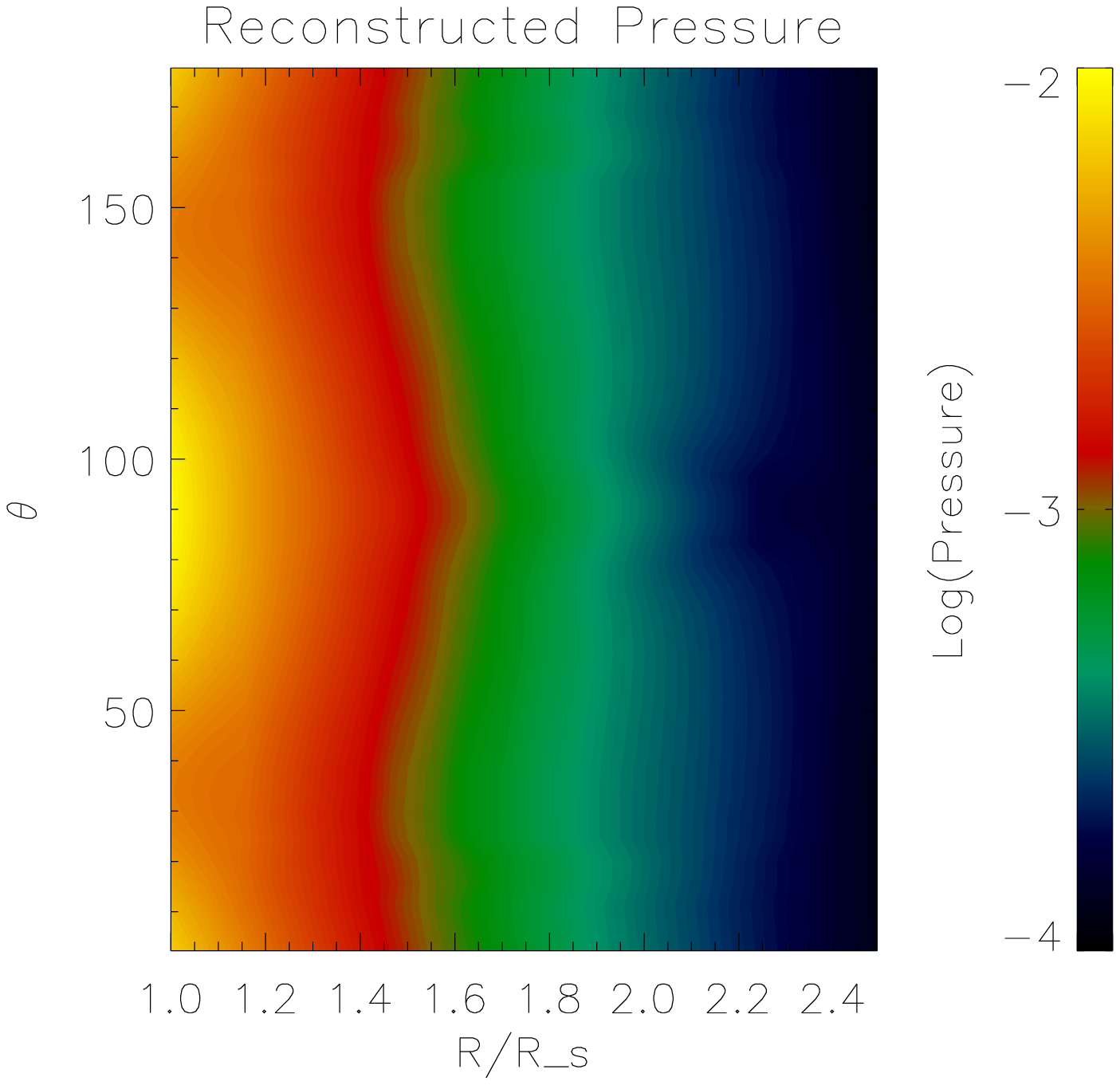}
\includegraphics[bb=0 0 410 410,,clip,height=3cm,width=4.0cm]{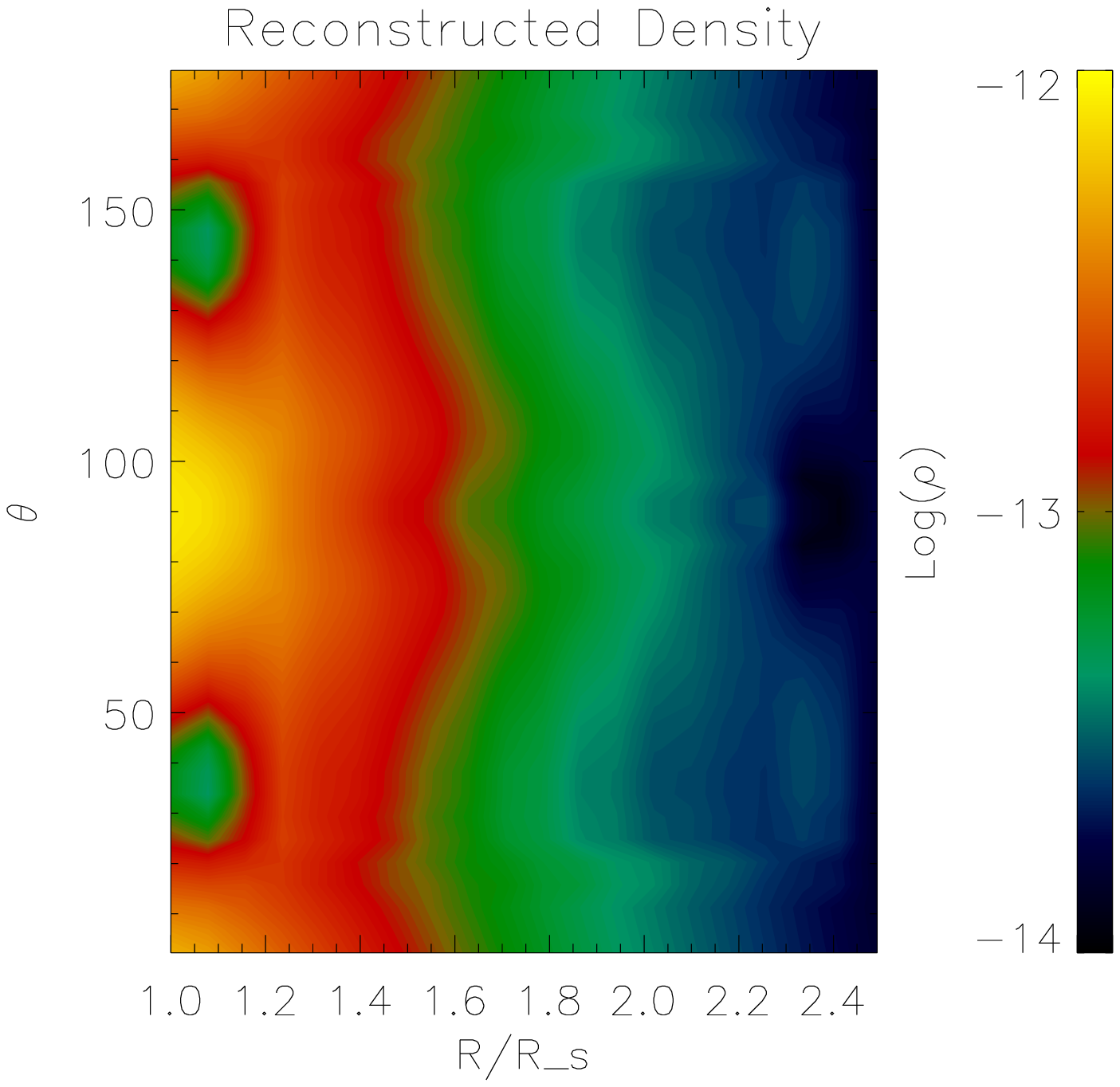}}
\caption{Plots of the spatial variation of the pressure $p$ (left column) and
density $\rho$ (right column) in logarithmic scaling in the $r$-$\theta$-plane for $\phi =0$.
The top, center and bottom panels correspond to the exact solution, the spherically symmetric
stratified atmosphere used as initial state for the iteration (also identical to the background atmosphere used in the
analytical solution) and the result of the numerical calculation. The main features of the analytical
solutions are clearly recovered by the numerical method, but some small differences are still present.
}
\label{fig2}
\end{figure}

\section{Results}
\label{results}
The numerical code based on the optimization method described in section \ref{equations} has been
run with initial conditions given by a potential field calculated from just the $B_r$ component of
the above exact solution on the boundaries.
The initial conditions for density and pressure are as in a stratified
atmosphere. This configuration balances the radial pressure
gradient and the gravity force. Consequently the
initial plasma is structured only in the radial direction and
$p$ and $\rho$ are invariant in $\theta$ and $\phi$. We use the exact
solution to prescribe the boundary conditions for ${\bf B}, p, \rho$.
Our code solves the magnetohydrostatic equations (\ref{1})-(\ref{2})
in the computational box with respect to these boundary conditions. To evaluate
the performance of our code we compare the result with the exact solution.

A visual impression of the exact magnetic field is given in the top panel of Fig. \ref{fig1}. For
comparison the potential field used as initial condition for the iteration is shown in the middle
panel of Fig. \ref{fig1}. One notices in particular that the potential field has a different
connectivity than the analytical MHS solution. The numerical method will thus have to change
the magnetic field connectivity during the iteration process.

In Fig. \ref{fig2} we show $r$-$\theta$ cuts of pressure (left) and density  (right) for the
angle $\phi=0$. The top panels show these quantities for the exact solution, the middle panels
for the initial condition and the bottom panels show the result of the numerical calculation.
Visual inspection of these figures gives the impression that the method does achieve a good,
but not perfect agreement with the exact solutions. Especially the regions close to the $\theta$
boundaries still show noticeable differences, which may be caused by the general problems with
convergence closer to the poles due to the deformation of the numerical grid using spherical
coordinates.  A more sophisticated numerical grid, as discussed in section \ref{conclusions},
will probably help to overcome these problems.

\begin{figure}
\includegraphics[height=6cm,width=8cm]{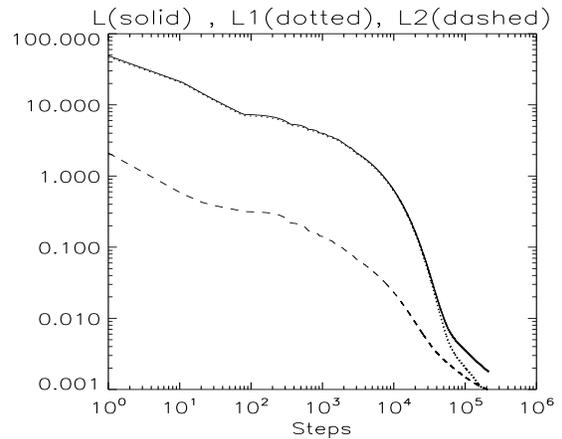}
\caption{Evolution of the functionals $L$ (solid line), $L_1$ (dotted line) and $L_2$ (dashed line; for definitions see text) during the iteration process.
All three functionals decrease by several orders of magnitude during the iteration.
}
\label{fig3}
\end{figure}

To check the quality of the reconstruction in a more quantitative way we use a number of methods.
The results are presented in table \ref{table1}.
First, we evaluate how well the force balance condition
and the solenoidal condition are fulfilled. To do this we evaluate
\begin{itemize}
\item the functional
 $L$ as defined in (\ref{defL_Bprho}),
\item the functional
$L_1=\int \frac{\left|(\nabla \times {\bf B}) \times {\bf B}
-\mu_0 \, \nabla p -\mu_0 \, \nabla \Psi  \, \rho  \right|^2}{B^2}
\, r^2 \, \sin \theta \, dr \, d \theta \, d \phi$, telling us how well the force balance condition
is satisfied,
\item the functional
$L_2=\int |\nabla \cdot {\bf B}|^2 \, r^2 \, \sin \theta \, dr \, d \theta \, d \phi$, which tells us how well
the solenoidal condition is satisfied.
\end{itemize}
The evolution of the functionals $L, \, L_1, \, L_2$
during the numerical computation is shown in figure \ref{fig3} in a double logarithmic plot.
The functionals $L$ and $L_1$ decrease rapidly by about five orders of magnitude. The functional $L_2$,
which represents the solenoidal condition decreases by about three orders of magnitude. The entries in
the table give the values of the functionals at the end of the numerical calculation.

Following \citet{amari:etal06} and \citet{wiegelmann:etal06b} we also provide
the infinity norms
\begin{itemize}
\item $\parallel \nabla \cdot {\bf B} \parallel_{\infty}$
\item $\parallel {\bf j } \times {\bf B}-\nabla p -\rho \nabla \Psi \parallel_{\infty}$,
\end{itemize}
which are defined as the supremum of the divergence and the total force density. For comparison,
we also provide these values for the exact analytic solution if used in a discretized form.
This allows us to estimate the discretisation error introduced by calculating the solution on a
finite numerical grid (Here $20,40,80$ grid-points in the $r, \theta, \phi$ direction, respectively).

We compare the reconstructed equilibrium directly with the known analytic equilibrium
by a number of quantitative measures as defined by
\citet[][see section 4 of that paper]{schrijver:etal06} . The figures quantify the
difference between two discretized vector fields ${\bf B}$ (known analytic field evaluated on the numerical grid) and ${\bf
b}$ (reconstructed field). These measures are:
\begin{itemize}
\item the vector correlation
$C_{\rm vec}=  \sum_i {\bf B_i} \cdot {\bf b_i}/ \left( \sum_i |{\bf B_i}|^2 \sum_i
|{\bf b_i}|^2 \right)^{1/2}$,
\item the Cauchy-Schwarz inequality
$C_{\rm CS} = \frac{1}{N} \sum_i \frac{{\bf B_i} \cdot {\bf b_i}} {|{\bf B_i}||{\bf
b_i}|}$,
where $N$ is the number of vectors in the field;
\item the normalized vector error
$E_{\rm N} = \sum_i |{\bf b_i}-{\bf B_i}|/ \sum_i |{\bf B_i}|$,
\item the mean vector error
$E_{\rm M} = \frac{1}{N} \sum_i \frac{|{\bf b_i}-{\bf B_i}|}{|{\bf B_i}|}$
\item the magnetic energy of the reconstructed field
divided by the energy of the analytical field
$\epsilon = \frac{\sum_i |{\bf b_i}|^2}{\sum_i |{\bf B_i}|^2}$.
\end{itemize}
The quality of the reconstruction of the
pressure $p$ and the density $\rho$ is quantitatively assessed
by correlating the analytic and the reconstructed solutions using
the linear Pearson correlation coefficients  (called
correlation $p$ and correlation $\rho$ in table \ref{table1}).
Finally, we provide the number of iteration steps and computing time
for a single processor run on a common workstation.
\begin{table}
\caption{
The table provides several figures of merit which can be used to assess the
quality of the reconstructed solution.
We compute all figures for the complete computational domain.
The analytical reference field was specified in the cones
$11.25^\circ \leq \theta \leq 168.75^\circ$.
For the calculation presented here we used the following iteration parameters:
$\mu=1, \, \nu=0.01, \, \xi=1$.}
\label{table1}
\begin{tabular}{c|lll}     
\hline
 & Ref. & Potential & Reconstruction \\
$L$   &$0.003$&$0.004$&$0.002$\\
$L_1$ &$0.001$&$0.002$&$0.001$\\
$L_2$ &$0.002$&$0.002$&$0.001$\\
$\parallel \nabla \cdot {\bf B} \parallel_{\infty}$&$0.448$&$0.582$&$0.448$ \\
$\parallel {\bf j } \times {\bf B}-\nabla p -\rho \nabla \Psi \parallel_{\infty}$
&$0.137$&$0.186$&$0.143$ \\
$C_{\rm vec}$&$1$&$0.946$&$0.9997$ \\
$C_{\rm CS}$&$1$&$0.810$&$0.997$ \\
$E_{N}$&$0$&$0.378$&$0.021$ \\
$E_{M}$&$0$&$0.524$&$0.042$ \\
$\epsilon$&$1$&$0.951$&$1.0008$ \\
Correlation $p$&$1$&$0.975$&$0.9998$ \\
Correlation $\rho$&$1$&$0.928$&$0.9990$ \\
No.\ of Steps &&&$214220$\\
Computing time &&&$12h27min$ \\
\hline
\end{tabular}
\end{table}
\section{Conclusions and Outlook}
\label{conclusions}
\begin{figure*}
\includegraphics[angle=-90, width=\textwidth]{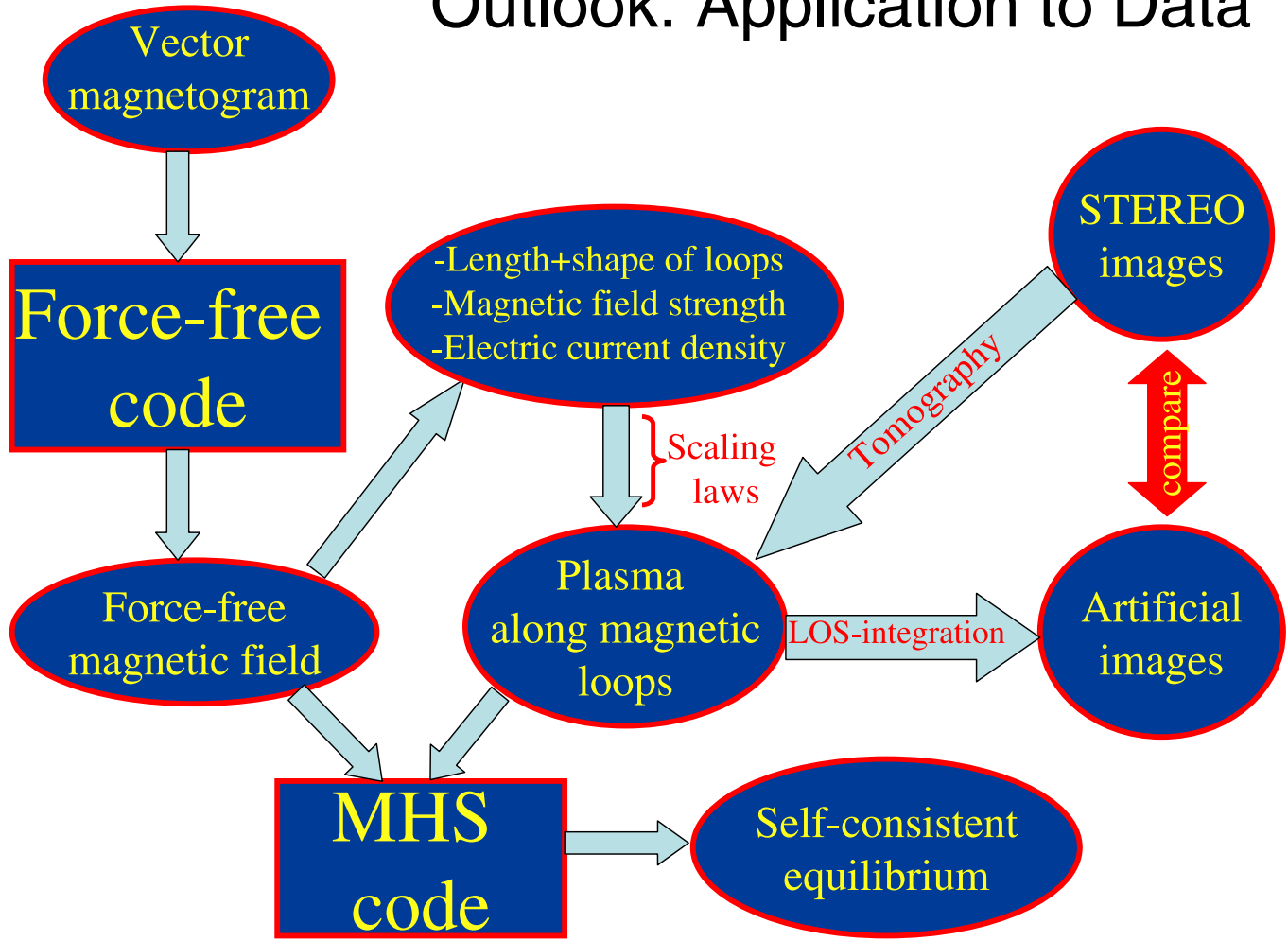}
\caption{Outlook: How can the tool become applied to data? The basic idea is to compute first a nonlinear
force-free field from observed vector magnetograms and then model the plasma along the loops. Our newly developed
magneto-hydrostatic code uses the resulting magnetic field and plasma configuration as input for the computation
of a self-consistent equilibrium.}
\label{fig4}
\end{figure*}
We have extended the optimization method originally proposed for the reconstruction of force-free magnetic fields
\citep{wheatland:etal00} to global magnetohydrostatic equilibria including the pressure force and the gravitational
force in spherical geometry. The proposed generalization of the optimization method leads to two additional equations
for the pressure and the density that have to be solved simultaneously to the magnetic field equation. Boundary
conditions for the magnetic field, the pressure and the density are necessary to complete the problem.

We have implemented a numerical code based on the proposed method and have tested the code using a known
three-dimensional magnetohydrostatic equilibrium \citep{neukirch95}. The numerical calculation is started
from a potential field with the same radial magnetic field component as the analytical equilibrium
on the boundary. The initial pressure
and density distribution are a spherically symmetric stratified atmosphere in hydrostatic balance. Both visual
inspection of the results as well as a quantitative analysis using various diagnostic measures indicate that
the method works well and converges to the analytic equilibrium.
For the presented tests we used a low spatial resolution and got a relatively long computing
time (about 200 000 iteration steps) until convergence. In experiments with the force-free version
of our spherical code \citep[see][]{wiegelmann07}
we found that the computing time scales with $N^{5.4}$ regarding the number
of grid points $N$ in one spatial direction. This is somewhat slower as the theoretical
estimate of $N^5$ for a cartesian optimization code obtained by \citet{wheatland:etal00}.
The spherical magnetohydrostatic code is significantly slower than the cartesian
force-free code for two reasons.
\begin{enumerate}
\item  The convergence of the numerical grid towards the poles requires a
sufficiently small time-step.
\item The plasma $\beta$ might vary strongly in the entire region and in particular
low-$\beta$-regions require very small time-steps to compute the magnetic field and
plasma simultaneously, because small changes in the Lorentz force can result in
considerably large changes in the low $\beta$ plasma.
\end{enumerate}
Point 1.) can be addressed by using a more sophisticated numerical grid, e.g. the
Yin-Yang grid developed by \citet{kageyama:etal04},
which has been applied in geophysics \citep[see e.g.][]{yoshida:etal04}.
This overset grid contains two complementary grids which lead to an almost
uniformly spaced spherical grid.
An additional advantage of the Yin-Yang grid is that it is suitable for massive
parallelization. The Yin-Yang grid has been applied for geophysical simulations
 on the Earth simulator super-computer in Yokohama.
To speed up the 2. point one might compute first the magnetic field alone as a nonlinear
force-free field (which is a reasonable approximation in low-$\beta$-regions) and switch on
the self-consistent plasma iteration only after for a fine-tuning.
A multi-scale approach, as recently implemented to speed up our force-free cartesian
optimization code \citep[see][for details]{metcalf:etal07} is also an option worth trying
for the spherical implementation of our force-free and magnetohydrostatic codes.
We are confident, that the above mentioned potential for improvements together with a
massive parallelization will allow us to apply our newly developed method to real data with
a reasonable grid resolution.

In figure \ref{fig4} we outline a scheme on how the code might be applied to data.
{\tw As boundary conditions on pressure and density are not directly measured, we propose
the following approach.}
The basic idea is to compute first a force-free magnetic field and then model
the plasma along the magnetic loops, e.g., by the use of scaling laws and optionally
with the help of a tomography code. Such an approach has been used by
\citet{schrijver:etal04} by using a global potential magnetic field and specifying
free scaling law parameters (e.g., heating function) by comparing artificial plasma
images (created by line-of-sight integration from the model plasma)
with X-ray and EUV observations. We propose
to generalize this approach by using a nonlinear force-free magnetic field model
and compare the model plasma with observations from two viewpoints as provided
by the STEREO-mission. Optional STEREO-images can be used additionally to approximate the
coronal density distribution by a tomographic inversion.
As a consequence of this step (modelling the plasma along a magnetic loop) the plasma pressure
is consistent along the loops. Different values for the pressure on different
field-lines will violate the force-free condition for the magnetic field, however,
and the configuration is not exactly in a magnetohydrostatic equilibrium.
Finite pressure gradients have to be compensated by a Lorentz force. This computation
can be done with the help of the program described in this paper.
We propose to use the force-free magnetic field configuration and the model plasma as
initial state for our newly developed magnetohydrostatic code to compute a self-consistent
MHS-equilibrium. For a low $\beta$ plasma the back-reaction of the plasma onto the magnetic
field will be small, for higher values of $\beta$ (as found e.g., in helmet streamers) the
magnetic field might change significantly. As a result of this approach one has reconstructed
the 3D coronal magnetic field and plasma configuration self-consistently within the
magnetohydrostatic approach and the model is consistent with measured photospheric vector
magnetograms and observed coronal images from different viewpoints as well.

\begin{acknowledgements}
The work of T. Wiegelmann was supported by DLR-grant 50 OC 0501. T. Neukirch acknowledges financial support by STFC.
The work of P. Ruan was supported by the International Max-Planck Research School on Physical Processes in the
Solar System and Beyond at the Universities of Braunschweig and Goettingen.
Financial support by the European Commission through the SOLAIRE Network (MTRN-CT-2006-035484) is also gratefully acknowledged.
We would like to thank the referee, Mike Wheatland, for useful remarks to improve this paper.
\end{acknowledgements}
\bibliographystyle{aa}
%

\end{document}